# Subdiffraction and spatial filtering due to periodic spatial modulation of the gain/loss profile

K. Staliunas[1,2] R. Herrero[2], and R.Vilaseca[2]

[1]*Institució Catalana de Reserca i Estudis Avançats (ICREA), Pg. Lluis Companys, 23, 08010 Barcelona, Spain, EU*

[2]*Departament de Fisica i Enginyeria Nuclear, Universitat Politècnica de Catalunya, Colom 11, 08222 Terrassa, Barcelona, Spain, EU*

**Abstract:**

We investigate light propagation through materials with periodically modulated gain/loss profile in both transverse and longitudinal directions, i.e. in material with two-dimensional modulation in space. We predict effects of self-collimation (diffraction-free propagation) of the beams, as well as superdiffusion (spatial frequency filtering) of the beams in depending on the geometry of the gain/loss lattice, and justify the predictions by numerical simulations of the paraxial wave propagation equations.

It is well known, that the materials with the refractive index modulated in space on the wavelength scale, i.e. the so-called photonic crystals (PCs), bring about a significant modification of the propagation properties of waves, both in time and space domains. In time domain the usual (temporal) dispersion is modified, and photonic band gaps appear in the frequency spectra [1,2]. The photonic bandgaps are, perhaps, the most celebrated property of photonic crystals. In the space domain, the spatial dispersion (diffraction) can also be modified. This leads, analogously to the temporal case, to the appearance of bandgaps in terms of the propagation direction (i.e. in the propagation wavevector domain). The character of diffraction in the propagation bands is also modified: it can result in negative diffraction leading to lensing [3], or can result to the zero-diffraction leading to self-collimation [4] (also called diffractionless or subdiffractive propagation [5,6]).

Whereas the PCs, the materials with spatially modulated refractive index, are the subject of intensive study, the seemingly analogous materials, those with gain/loss modulation (GLM), do not enjoy a sensible interest. One reason for such "discrimination" is that the GLM materials do not possess the celebrated property of the PCs – the bandgaps. If the refractive index modulation pushes the frequencies of the harmonic field components one from another, and thus opens bandgaps at around their cross-points, the modulation of the gain/loss does the opposite – it closes the bandgaps. This can be illustrated on a simple example of two coupled spatial harmonic field components (equivalently for two coupled oscillators oscillating and evolving in time):

$$\frac{dA_1}{dz} = ik_1 A_1 + \gamma A_2 \tag{1.a}$$

$$\frac{dA_2}{dz} = ik_2 A_2 + \gamma A_1 \tag{1.b}$$

Here $k_{1,2}$ are the propagation wavenumbers of the spatial field harmonics, and $\gamma$ is the coefficient of coupling due to a periodic spatial modulation in the material. The refractive index modulation results in reactive coupling ($\gamma = im$ with $m$ real-valued), and the modulation of gain/loss results in dissipative coupling ($\gamma = m$). A simple analysis reveals substantially different properties of the eigenmodes $A_{1,2} \propto Exp(iKz)$ of Eqs (1) in the reactive and diffusive cases. In the reactive case the dispersion curves of the harmonic components push one another, and a band-gap appears at around the degeneracy of the propagation wavenumbers $k_1 = k_2$ (analogously the bandgap appears at the degeneracy of frequency $\omega_1 = \omega_2$ for the case of two coupled temporal oscillators). In contrast, in the dissipative coupling case, the dispersion curves "pull" one another and their propagation wavenumbers (their frequencies in the case of temporal oscillators) tend to lock at some interval around the degeneracy. The solution of (1):

$$K_{1,2} = (k_2 + k_1)/2 \pm \sqrt{((k_2 - k_1)/2)^2 - \gamma^2} \tag{2}$$

is represented in Fig.1 for the reactive and dissipative couplings, demonstrating the different behaviors in these two cases.

In this way, the bandgaps do not appear in the GLM systems. We show in this letter, however, that the modification of the spatial dispersion given by Eq.(2) can lead to a variety of nontrivial spatial effects. Indeed the spatial propagation effects in PCs appear due to the deformation (flattening) of the interacting (mutually pushing) spatial dispersion curves. In the present case of GLM, the spatial dispersion curves also become substantially modified due to the mode pulling and locking. The novel spatial phenomena associated with these deformation of the spatial dispersion curves is the topic of the present paper.

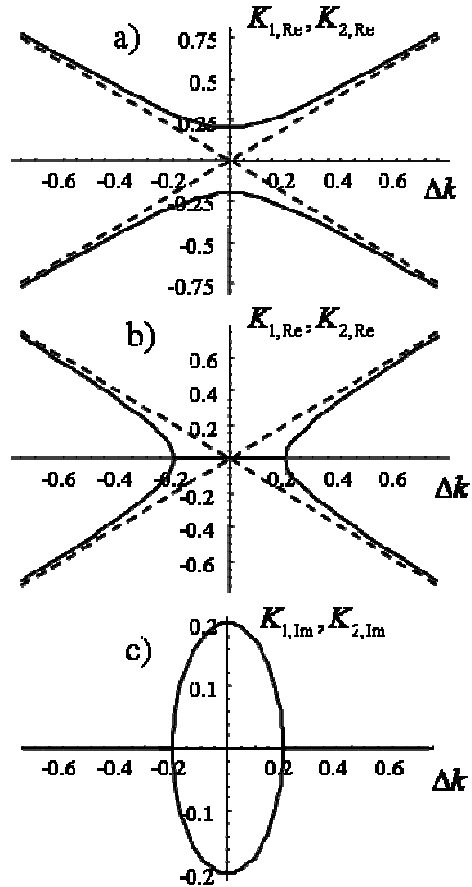

*Fig.1. a) Splitting of the propagation wavenumbers in the case of the reactive coupling of two spatial harmonics (due to the refractive index modulation) $\gamma = im$ around the degeneracy point $\Delta k = k_2 - k_1 \to 0$. b) In the case of dissipative coupling (due to gain/loss modulation) $\gamma = m$ the harmonics lock one to another instead of the mutually pushing. However, in the latter case the imaginary parts of the propagation wavenumbers (the growth exponents) are no more zero (c). In all cases $m = 0.2$.*

We note, that the system of two coupled field harmonics (1) used above to illustrate mode pushing and locking in Fig.1, corresponds to the case of transversal (1D) modulation of the material: the 1D modulation couples two field harmonics, and results into two relevant Bloch modes. The case of 1D modulation in one transverse dimension, as well as 2D modulation in two transverse dimensions (but no modulation in the longitudinal direction), have been recently investigated in [7]. We note that the novel results reported in the present letter lie on the modulation *along* as well as perpendicular to the propagation direction of light. Such configuration requires expansion at least into three harmonics, similarly to the approach used recently in the case of PCs in [6].

We first study analytically the situation described above in the simplest approach of three interacting field harmonics. We find two basic regimes: 1) a subdiffractive regime, where flat segments on the spatial dispersion curve appear and also no amplification takes place for the corresponding transverse wave vectors, resulting in a process analogous to self-collimation; and 2) a spatial filtering regime, where a highly directional angular gain profile occurs with a nearly flat dispersion curve. We prove these analytical predictions by numerical integration of the corresponding propagation equations in the paraxial approximation.

In analogy with the previous studies of light propagation in PCs [6], we use the paraxial approximation: The gain/loss modulation (GLM) function is $g(x,z) = 4g_0 \cos(q_\perp x)\cos(q_\parallel z)$, which represents a spatially-periodic lattice with transverse and longitudinal modulation wavenumbers $(q_\perp, q_\parallel)$. Gain (loss) occurs in areas $g(x,z) > 0$ ($g(x,z) < 0$); the average the gain is considered to be zero, without the loss of generality. The normalized paraxial propagation equation reads:

$$\frac{\partial A}{\partial Z} = \left( i \frac{\partial^2}{\partial X^2} + V(X,Z) \right) A \qquad (3)$$

where $A(X,Z)$ is the slowly-varying envelope of the electromagnetic field. The normalizations are adapted from [6], where $X = xq_\perp$ is the normalized transverse coordinate; $Z = zq_\perp^2/2k_0$ is the normalized longitudinal coordinate ($k_0 = 2\pi/\lambda$ is the wavenumber of the incident monochromatic light), and

$$V(X,Z) = 4m \cdot \cos(X)\cos(Q_\parallel Z) \qquad (4)$$

where $Q_\parallel = 2q_\parallel k_0 / q_\perp^2$ is the normalized longitudinal component of the wavevector of the modulation, and $m$ is the normalized modulation parameter $m = g_0 k_0 / 4q_\perp^2$. The modulation is considered harmonic without losing the generality. We note that the GLM profile averaged with respect to the longitudinal direction (i.e. integrated over one transverse modulation period) is homogeneous, therefore no trapping (Peierls-Nabarro) effects are to be expected.

In order to be able to perform some analytical predictions we introduce a harmonic expansion in terms of the periodicity of the gain/loss profile:

$$A(X,Z) = \exp(iK_\perp X) \cdot$$
$$[a_0(Z) + a_{-1}(Z)\exp(-iX - iQ_\parallel Z) + a_1(Z)\exp(+iX - iQ_\parallel Z_\parallel) + ...] \quad (5)$$

Which, inserted into (3), (4), and after truncation to these three harmonics, results into:

$$da_0/dZ = -iK_\perp^2 a_0 + m(a_{-1} + a_{-1}) \quad (6.a)$$
$$da_{-1}/dZ = -i(K_\perp - 1)^2 a_{-1} + iQ_\parallel a_{-1} + ma_0 \quad (6.b)$$
$$da_1/dZ = -i(K_\perp + 1)^2 a_1 + iQ_\parallel a_1 + ma_0 \quad (6.c)$$

A similar equation system was used for the study of the self-collimation in PCs [6], except for the coupling coefficients, which are real valued now.

We perform next a standard analysis of Eqs. (6) by looking for the exponentially growing eigenmodes: $a_0, a_{-1}, a_1 \propto \exp(iK_\parallel z)$, where the exponential growth is incorporated into the imaginary part of $K_\parallel$. The analytic expressions for the propagation exponents are cumbersome (they are roots of a third order polynomial), therefore the main analysis (diagonalization of the corresponding matrix) has been performed numerically.

*Sell-collimating regime:* Self-collimation occurs for $Q_\parallel > 1$, in a parameter range, which is "above" the triple-cross-point (the triple-cross-point is at $Q_\parallel = 1$, where the three parabolas, i.e. the three dispersion curves of the spatial harmonics (5) (Fig.2.a) cross at one point. This is in contrast with PCs case, where the condition $Q_\parallel < 1$ holds for the self-collimation regime [6]. By choosing a particular amplitude $m$ for the GLM, one can achieve the flattening of the dispersion curve for a given geometry, as shown in

Fig.2.b. This entails self-collimation. The behavior of the imaginary part of the propagation wavenumber of the Bloch mode is shown in Fig.2.c. Whereas the self-collimated modes propagate with a net gain/loss equal to zero, the modes propagating at some angles (those around the values corresponding to the crossing of the dispersion curves of two harmonic components) mutually lock and amplify. This means, that the self-collimated beam can develop instability with respect to the sidebands during the propagation.

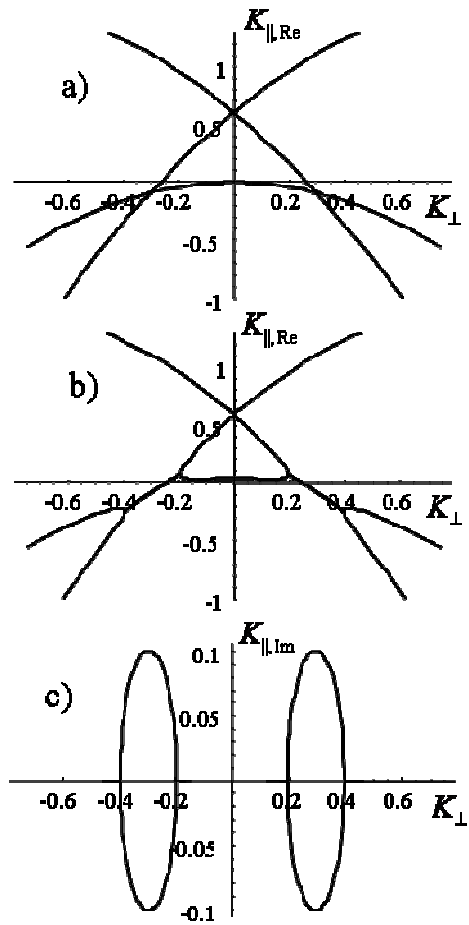

*Fig.2.* Propagation eigenvalues (real parts of $K_\parallel$) (a,b), and growth exponents (imaginary parts of $K_\parallel$) (c) for the three-coupled-mode model (6), in the self-collimation regime for $Q_\parallel = 1.6$. (a) illustrates the geometry of the dispersion parabolas of harmonic components with vanishing coupling ($m \to 0$). (b) shows the development of a self-collimation plateau on the dispersion curves of the Bloch modes for $m = 0.1$.

Analytic estimations of the width of the locking regions use two mode expansions separately in the vicinities of their corresponding cross point. We find that, for instance, the locking region at the right in Fig.2b,c) is given by the following expression:

$$\frac{Q_\| - 1}{2} - m < K_\perp < \frac{Q_\| - 1}{2} + m \tag{7.a}$$

Thus the width of the locked areas is $\Delta K_{lock} = 2m$. On the other hand the condition for self-collimation (flattening of the dispersion curve) can be obtained from a series expansion around the optical axis $K_\perp = 0$:

$$K_\|(K_\perp) = \frac{2m^2}{(Q_\| - 1)} + K_\perp^2 \left( \frac{8m^2}{(Q_\| - 1)^3} - 1 \right) + K_\perp^4 \frac{32m^2}{(Q_\| - 1)^5} + \ldots \tag{7.b}$$

Which leads to the condition of zero diffraction point: $Q_\| - 1 \approx 2m^{2/3}$. Eqs (7.a,b) allow also to estimate the width of the plateau defining the self-collimation region (Fig.2b): $\Delta K_{s-coll} \approx 2(m^{2/3} - m)$.

The predicted self-collimation can be confirmed by numerically solving the paraxial equation (3,4). Fig.3.a shows that the beam, upon entering the modulated material propagates without diffraction. However two sidebands grow, and after some propagation distance they become dominating (note in Fig.3.a that the transverse distributions of the field are normalized to its peak value at every propagation distance in so that actually there is no attenuation of the central part of the beam). The nondiffractive character of the central part of the beam is demonstrated in Fig.3.c, which compares the width of the central, self-collimated, part of the beam with the width of an analogous beam propagating in a homogeneous material. The small oscillations appear because of the influence of the amplifying sidebands. And finally, the angular distribution of the gain is presented in Fig.3.d, as obtained by numerical integration of plane waves entering into the media of modulated gain/loss at varying angles.

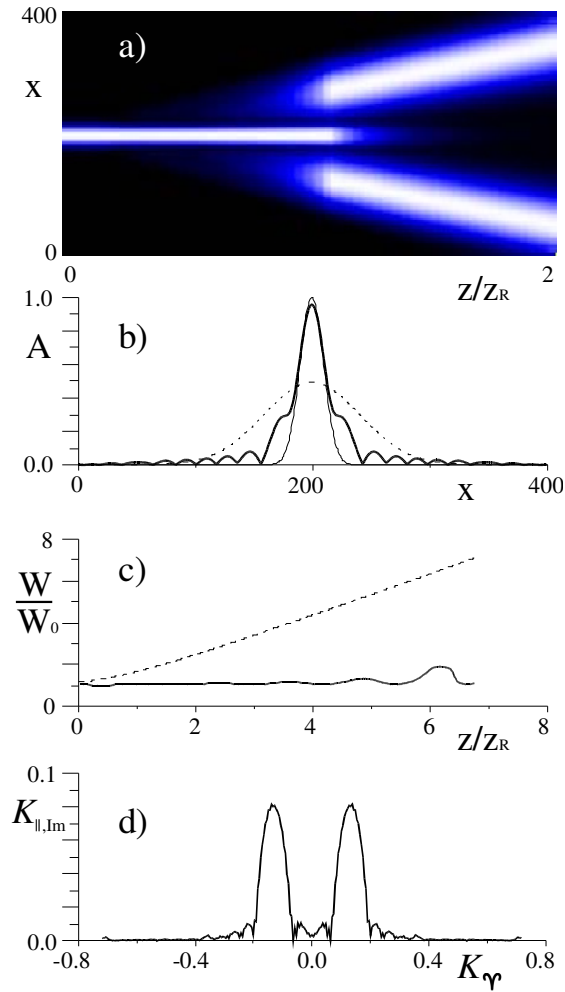

*Fig.3 a) Field amplitude for a propagating beam through a GLM material with m=0.1 and $Q_{\parallel}$=1.6. The peak amplitude in each transverse section has been normalized to unity. The initial width of the Gaussian beam is $W_0$=13.3 corresponds to a Rayleigh length $Z_R$=88.8. b) Profiles of the initial Gaussian beam (thin curve) and of the beam propagated a distance $Z=4Z_R$ along the modulated material and space filtered to remove the amplified transverse modes (bold curve). The profile of the beam propagating through a homogeneous material is shown for comparison (dashed curve). c) Width of the central part of the beam propagating along the modulated and a homogeneous material (dashed curve). The oscillations appear because of the influence of the amplifying sidebands. d) Imaginary part of the propagation wavenumber obtained from the propagation of plane waves through the material entering at varying angle.*

***Space-filtering regime:*** Close to the triple-cross-point of the dispersion parabolas, i.e. at around $Q_\parallel = 1$ the above described self-collimation is absent. The geometry of the dispersion curves of the harmonic components is given in Fig.4.a. Characteristic is that for a definite range of amplitudes of the gain/loss modulation (as shown in Fig.4.b and Fig.4.c) a narrow gain profile along the optical axis appears, which indicates a strong angular dependence of the gain around $K_\perp = 0$. The latter acts as a spatial filtering, as it results in a discrimination of the high transverse harmonics.

For analytical estimations we perform a series expansion of the eigenvalues considering the smallness conditions $(Q_\parallel - 1) = O(\varepsilon)$, and $K_\perp^2 = O(\varepsilon)$. We obtain for the amplified Bloch modes:

$$K_{\parallel, \text{Re}}(K_\perp) = \frac{Q_\parallel - 1}{m} - \left(1 + \frac{Q_\parallel - 1}{m^2}\right) K_\perp^2 + \ldots \tag{8.a}$$

$$K_{\parallel, \text{Im}}(K_\perp) = \frac{\sqrt{2}}{m}\left(m^2 - K_\perp^2\right) + \ldots \tag{8.b}$$

Eqs (8) reveals that the maximal amplification is $K_{\parallel, \text{Re,max}} = \sqrt{2}m$, and the half-width of the amplification line is $|\Delta K_\perp| = m$. Also we find that the dispersion curve becomes flat at: $m^2 = 1 - Q_\parallel$; The latter also means that for $Q_\parallel > 1 - m^2$ the diffraction is normal, but for $Q_\parallel < 1 - m^2$ the diffraction is anomalous (negative). Physically this means that the amplified beam attains a negative curvature of the phase front, which leads to beam focusing behind the medium.

The eigenvector corresponding to the amplified mode can be expanded at around the optical axis $K_\perp = 0$:

$$\left\{\frac{1}{\sqrt{2}} + \frac{i(Q_\parallel - 1)}{4m} + K_\perp\left(\frac{i}{m} - \frac{(Q_\parallel - 1)}{\sqrt{2}m^2}\right) + \ldots, \; 1, \; \frac{1}{\sqrt{2}} + \frac{i(Q_\parallel - 1)}{4m} - K_\perp\left(\frac{i}{m} - \frac{(Q_\parallel - 1)}{\sqrt{2}m^2}\right) + \ldots\right\} \tag{9}$$

At $K_\perp = 0$ and in the limit $Q_\parallel = 1$ (triple cross point) the eigenvector is: $\left\{\frac{1}{\sqrt{2}}, 1, \frac{1}{\sqrt{2}}\right\}$, which means that in the output half of the energy of the amplified and filtered radiation goes to the central diffraction maximum, whereas the other half goes to the first diffraction components.

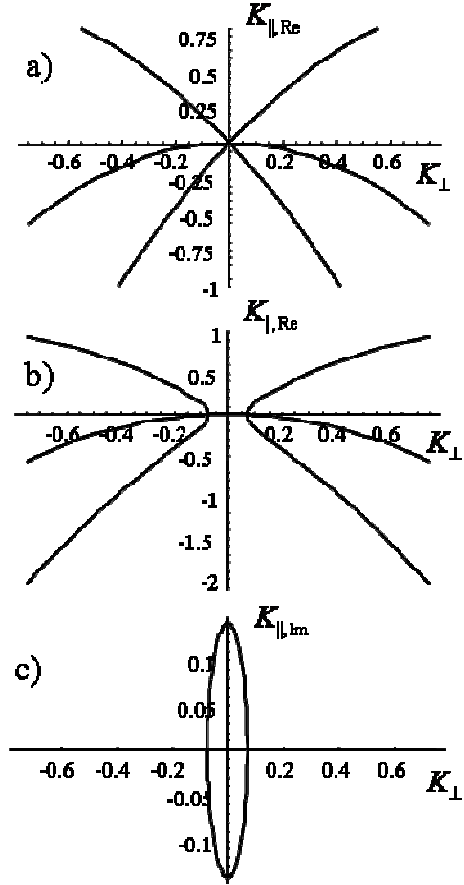

**Fig.4.** *The propagation eigenvalues (real parts of $K_\parallel$) (a,b), and the growth exponents (imaginary parts of $K_\parallel$) (c), for the three-coupled-mode model (6), in the spatial filtering regime. (a) Geometry of dispersion parabolas of harmonic components. (b) Plateau of mode-locking for $m = 0.1$ (c) Angular gain profile. $Q_\parallel = 1.0$*

We prove numerically with the paraxial model (3.4) the spatial filtering effect based on the above discussed angular selectivity of the amplification. Fig.5.a shows the evolution of the profile of the amplifying beam, exhibiting its diffusive character of broadening. We note that the diffusive broadening follows a $W \propto Z^{1/2}$ dependence in contrast to the diffractively spreading beams obeying $W \propto Z^1$. This dependence is shown in Fig.5.b.

The angular distribution of the gain is presented in Fig.5.c, obtained by numerical integration of plane waves entering into the media of modulated gain/loss at varying angles. The narrow angular spectrum of amplification corresponds well with that calculated semianalytically in Fig.4.

The negative diffraction obtained for $Q_\parallel < 1 - m^2$ values results in a beam focalization behind the GLM material. Fig.5d shows the beam propagation through a piece of GLM material of length $0.84Z_R$ resulting to the beam focalization at some distance behind the material. The detailed investigation of the beam focalization behind the periodic GLM material is to be studied separately.

The energy distribution among the central mode and sidebands at the crystal output from numeric simulations is in good agreement with the linear analysis. For this particular case (Fig.5d) the central beam energy is about 48.8% of the total energy.

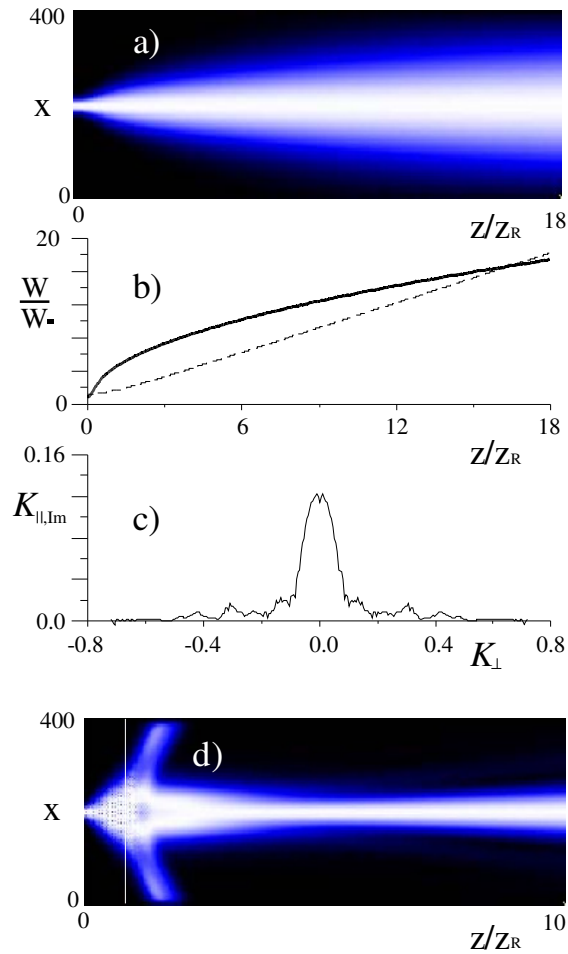

**Fig.5** *a) Propagation of a Gaussian beam through gain/loss modulated medium with $m=0.1$ and $Q_{||}=1.0$. The initial beam is as in Fig.3. b) The width of the beam propagated through the modulated medium (thick line) and homogeneous material (dashed line). c) Imaginary part of the propagation wavenumber obtained from the propagation of plane waves through the material in different directions. d) Focalization of the beam behind the for $m=0.1$ and $Q_{||}=0.85$. white vertical line indicates the rear face of the crystal.*

Concluding, we have predicted and demonstrated theoretically (by means of three-mode semianalytical expansion and by numerical integration of paraxial model) the existence of two novel phenomena of the propagation of the beams in media with gain/loss modulation in 2D (one longitudinal and one transverse direction). These are the self collimation, similar to that predicted in PCs, and spatial filtering, which has no analog in PCs. Seemingly the second phenomenon, spatial filtering, is of a technological importance. Self-collimation as shown above, is obscured by the fact that the amplification of the modes at some angles to the optical axis occur, which means the modulation instability of the beam propagating in self-collimated regime. The second predicted effect, the spatial-filtering, should allow preparing highly directional beams. In addition, the amplified beam can be automatically adjusted to focus at a definite distance by fine tuning of the modulation geometry around the triple-cross-point of the parabolas. As described above, one can achieve weak positive as well as a weak negative diffraction in the material. This is an advantage in applications, where one can not only prepare, and not only amplify, a highly directed beam, but additionally one can have additional focusing of that beam behind the modulated material (if working in slightly negative-diffraction regime).

How realistic, and how technologically relevant are the systems with modulated gain/loss? If the loss-modulated system is hardly attractive from the viewpoint of the applications (because of the irreversible loss of the energy), the modulated gain systems can have practical applications. Possible concrete systems could be optically pumped lasers and amplifiers, where the pump should be realized through a coherent interference pattern of at least three beams. Also an optical parametric oscillator pumped by at least three interfering non-collinear beams could result in effects similar to the ones studied here (a two beam interference could create a 1D lattice only). The concrete realization of the amplification schemes will be the object of separate study.

The presented analysis, which concerns a 2D case, could be also extended into the 3D case. The basic discussion on the crossing and locking of dispersion curves can be extended to the 3D dispersion surfaces. The essential difference with the 2D case is the symmetry of the lattice in the transverse plane, as explored e.g. in [8] in case of PCs. It comes out from a preliminary study that the different symmetries of the modulation in the transverse plane result in a different transverse distribution of the amplified wavenumbers in the spatial Fourier domain, e.g. a square lattice results in approximately square distributions, and the lattices of hexagonal or octagonal symmetry result in more round ones.

The work was financially supported by Spanish Ministerio de Educación y Ciencia and European Union FEDER through project FIS2005-07931-C03-03.